\documentclass[aip,rsi,reprint]{revtex4-1}
\usepackage{graphicx}
\usepackage{fontenc}

\begin{document}


\title{Embedded Control System for Mobile Atom Interferometers} 



\author{Bola Malek}
\author{Zachary Pagel}
\author{Xuejian Wu}
\affiliation{Department of Physics, University of California, Berkeley, California 94720, USA}
\author{Holger M\"uller}
\email[]{hm@berkeley.edu}
\affiliation{Department of Physics, University of California, Berkeley, California 94720, USA}
\affiliation{Molecular Biophysics and Integrated Bioimaging, Lawrence Berkeley National Laboratory, Berkeley, California 94720, USA}

\date{\today}
\begin{abstract}
Atom interferometers require precise control of digital, analog, and radio frequency signals for effective operation. In this paper, we propose and implement a control system for mobile atom interferometers. The system consists of a microcontroller and peripherals to synthesize radio frequency signals and to read or write analog signals. We use the system to operate a mobile atomic gravimeter by controlling 7 analog outputs, 16 digital outputs, 2 radio frequency channels, and 1 analog input. Our control system eliminates dead time between repetitions of the measurement and, consequently, improves the sampling rate of our atomic gravimeter by more than a factor of 2 and the sensitivity by more than a factor of $\sqrt{2}$ compared to the system based on a desktop computer.
\end{abstract}


\maketitle 

\section{Introduction}
Advancements in atomic physics, in particular atom interferometry, have led to the development of laboratory-based sensors with increased accuracy, sensitivity, and long term stability compared to their classical counterparts\cite{parker2018measurement, overstreet2018effective}. Engineering advances result in instruments with decreased size and increased portability\cite{rushton2014contributed, kitching2011atomic}. Examples include atomic clocks\cite{at_clock18, grotti2018geodesy}, gyroscopes\cite{muller2009compact, hoth2016point, rakholia2014dual}, and gravimeters\cite{fang2016metrology, freier2016mobile}. Atom interferometers were deployed aboard ships\cite{bidel2018absolute}, a drop tower\cite{muntinga2013interferometry}, airplanes\cite{barrett2016dual}, and the International Space Station\cite{williams2016quantum}. Further miniaturization of component subsystems are necessary for field applications.

These systems often rely on desktop computers for timing and control using software packages such as Cicero Word Generator, which provides a convenient graphical user interface (GUI) for iterative development of atomic physics experiments\cite{keshet2013distributed}. Some trapped atom experiments\cite{pruttivarasin2015compact, yu2018performance} use field-programmable gate arrays (FPGAs). While these systems have enabled many measurements, they are restrictive for use outside of the lab. Systems based on microcontrollers (MCUs), known as embedded systems, offer several advantages in field settings such as small size, low power consumption, and flexibility. Such systems have been used for locking in optics experiments\cite{huang2014microcontroller}, laser spectroscopy\cite{eyler2013instrumentation}, event sequencing in atomic physics\cite{eyler2011single}, and operating an atomic clock\cite{deng2012embedded}.

This paper describes an embedded control system that operates a mobile atom interferometer for measuring gravity. The system is composed of a microcontroller, a generic analog interface, and a direct digital synthesizer (DDS). Operation of the atomic gravimeter with this system results in an improvement of the cycle time and the sensitivity.

\section{System Architecture}
Atomic sensors prepare a system of atoms, coherently manipulate it, then probe it. This process requires sequences of signals to open and close laser shutters, to ramp and switch magnetic fields, to sweep frequencies that modulate lasers, and to communicate with detectors.
\begin{figure}
    \centering
    \includegraphics[width=\columnwidth]{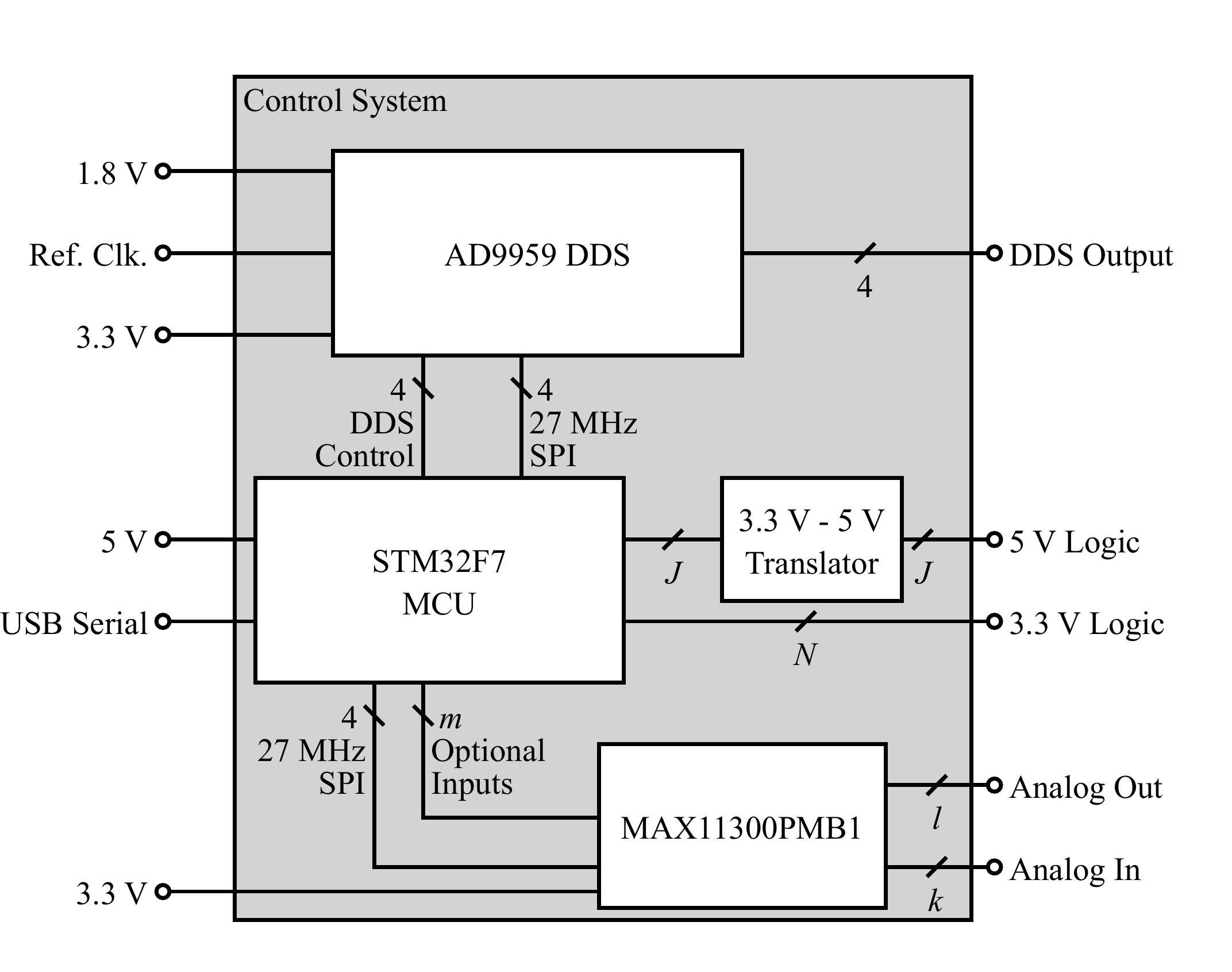}
    \caption{The system architecture. Lines with a slash indicate multi-wire connections. Certain connections allow for a variable number of wires as indicated by the variables: $m$, $N$, $l$, $J$, and $k$.}
    \label{fig:system}
\end{figure}
A sequence requires three types of control signals: digital, analog, and radio frequency (RF). Figure \ref{fig:system} shows the architecture of the control system, with each component dedicated to dealing with one type of signal. The MCU integrates the whole system by communicating with the other two boards. In addition, it handles all of the digital logic output. To incorporate functions not implemented on the MCU such as analog and RF control, the system employs the following two peripherals. The MAX11300PMB1 development board integrates the MAX11300, a chip from Maxim Integrated capable of generating and reading analog signals, with the necessary voltage regulators to power it from a 3.3 V source. This board serves as the analog front end for the system. The AD9959 Direct Digital Synthesizer (DDS) development board from Analog Devices provides 4 RF output channels that can be frequency, amplitude, or phase modulated.

To communicate between the MCU and the two peripherals, we use the Serial Peripheral Interface (SPI) protocol at 27 MHz. In addition, digital logic is used to further control the DDS modulation. A USB serial output is needed to reprogram the device and offload the data. The MCU includes 1 megabyte of flash on-chip for data storage if needed. For longer-term storage, a large memory peripheral can be accommodated in this architecture. The data is serialized and streamed over the USB serial interface. 

\subsection{Microcontroller}
An MCU was chosen for its low power consumption and fast development cycle compared to other types of controllers such as FPGAs. The STM32F7 MCU development board and its 217 MHz oscillator set the timing for every aspect of the system, except for the radio frequencies. By utilizing the cycle counter for timing control, delays can be controlled with the resolution of one instruction cycle, which is approximately 5 ns. The digital outputs of the MCU are organized into ports that can modified by a single instruction. On the STM32F7, ports have 16 digital channels each. When interfacing with an experiment,  all digital signals that change at the same time should be wired to the same port to ensure simultaneous changes. Note that other MCUs offer up to 32 or 64 outputs per port, and can be incorporated into this architecture.

The output logic level of the MCU is 3.3 V. To accommodate transistor-transistor logic (TTL), a level translator based on transistors or one on the MAX11300 (see page 28 of datasheet\cite{max_sheet}) can be used.

\subsection{Mixed-Signal Input-Output Control}
Detectors and voltage-controlled components such as voltage-controlled oscillators, attenuators, or current sources commonly operate over a range of 0 to 10 V. The voltage converters aboard the MAX11300PMB1 achieve this range. Other configurations allow for -5 to 5 V or -10 to 0 V ranges. Operational-amplifier circuits were used to accommodate other ranges while using the full (12-bit) resolution of on-board ADCs and DACs. Since experimental requirements tend to change during the development process, the MAX11300 was chosen for its ability to be reprogrammed and function as an analog-to-digital converter (ADC) and a digital-to-analog converter (DAC) on a variable number of channels.

The gravimeter requires that environmental parameters be slowly changed. In practice, this is realized by linear voltage ramps. The values of the steps in each ramp are precomputed then sent to the peripheral with delays to accommodate the settling time of the channel, which is 40 $\mu$s. Using this scheme, we can ramp multiple analog channels in parallel.

\subsection{DDS for Radio-Frequency Control}
RF modulation must be synchronized with other components in the system. As a result, the DDS board uses an external clock source. This clock can be between 10 and 30 MHz, and is be multiplied by a factor between 4 and 20 using a programmable oscillator and a phase-locking loop inside the DDS chip to set its clock. The MCU sets the parameters for frequency outputs and frequency sweeps using the SPI interface. The frequency is controlled using a 32-bit register, which allows for sub-mHz control. Frequency sweeps are set using the clock multiplier and the two parameters shown in figure 36 of the datasheet\cite{dds_sheet}: a frequency step and a time step. The time step is a 14-bit register, which allows for sub-$\mu$s resolution in most cases. A database of all possible frequency sweep rates is computed and maintained. When programming the control system, the desired rates are queries from this database and compiled with the program.

\subsection{Experimental Sequence}
An experimental sequence is a C++ program\footnote{If you are interested in the code, please contact us.} that is compiled then loaded to the MCU program memory. The program first initializes the MCU utilities needed and all the necessary peripheral drivers. Then, it sets the outputs to appropriate initial values and precomputes any values that will be frequently reused throughout the experimental sequence, specifically the analog ramp values. Finally, a loop contains the steps of the experiment. Arbitrary logic and real-time computation can be implemented in this loop. For example, the MCU can average, integrate, or filter ADC data to make real-time decisions, as opposed to relying on post-processing. This eliminates the need to save or transfer large amounts of raw, unprocessed data which often slows down experiments. The timing of each iteration is determined by the number of instructions in the body of the loop.

\section{Results}
\begin{figure*}
    \centering
    \includegraphics[width=\textwidth]{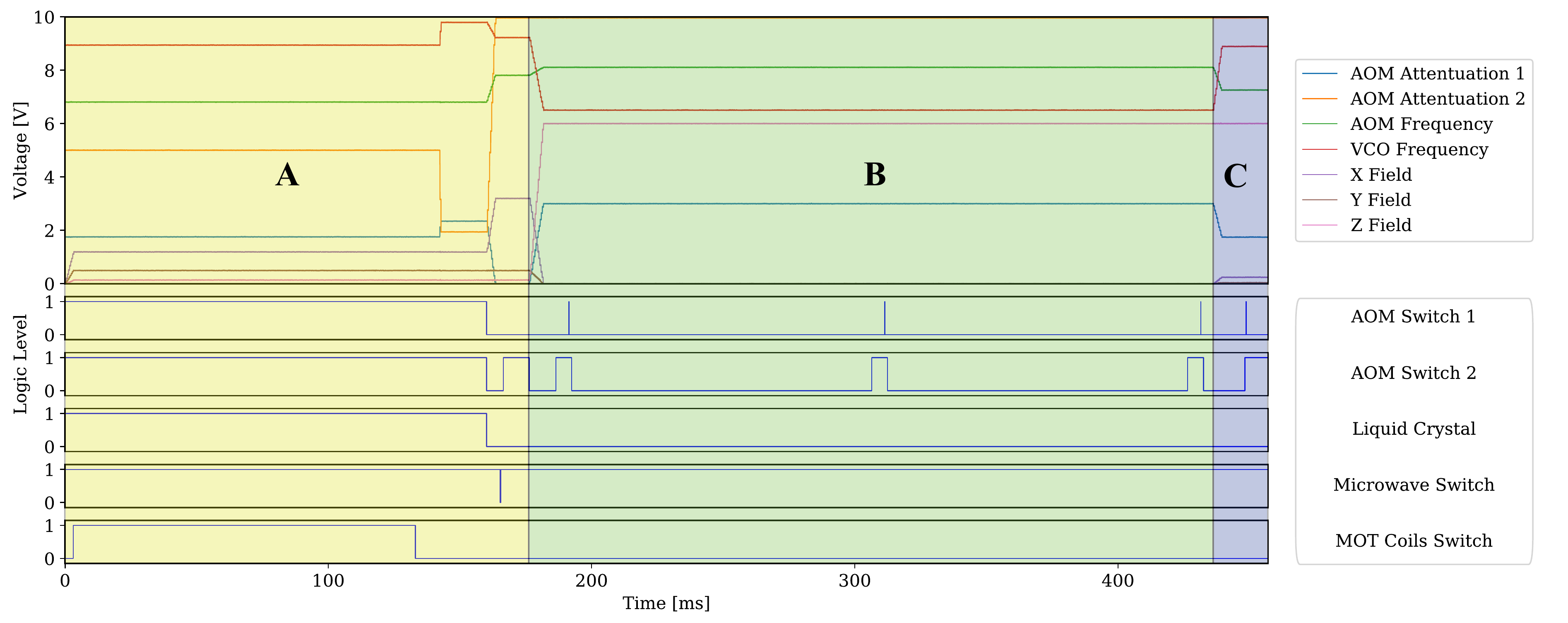}
    \caption{Measured timing diagram of select signals during the gravity measurement.These analog and digital signals are sampled using a Salaea Logic Pro 16 logic analyzer\cite{saleas_sheet} at 50  mega-samples per second (MSPS) and 500 MSPS respectively. A: atom cloud preparation, B: interferometry, and C: fluorescence detection.}
    \label{fig:timing}
\end{figure*}
Previously, we have developed a compact atom interferometer with a simple laser system based on a single diode laser and pyramid mirror magneto-optical trap (MOT) for inertial sensing\cite{wu2017multiaxis}. This atom interferometer uses Doppler-sensitive Raman pulses to interfere matter waves in a Mach-Zehnder geometry. The proportion of atoms in each state after recombination is detected by fluorescence imaging using a photodiode. Voltage-controlled current-sources, attenuators, and oscillators are used to control magnetic field strengths, laser powers, and laser frequencies. Accousto-optical modulators and electro-optical modulators allow for electrical control of the laser frequencies to access quantum transitions in the atomic system. RF signals linearly sweep the modulator frequencies to generate Doppler-sensitive laser pulses. Shutters and switches in the laser system, magnetic field driver, and microwave frequency subsystem are controlled using digital signals. The original system was controlled by system based on a desktop computer and the fastest achievable cycle time was more than 0.8 s longer than the specified time of the sequence.
\begin{figure}
    \centering
    \includegraphics[width=\columnwidth]{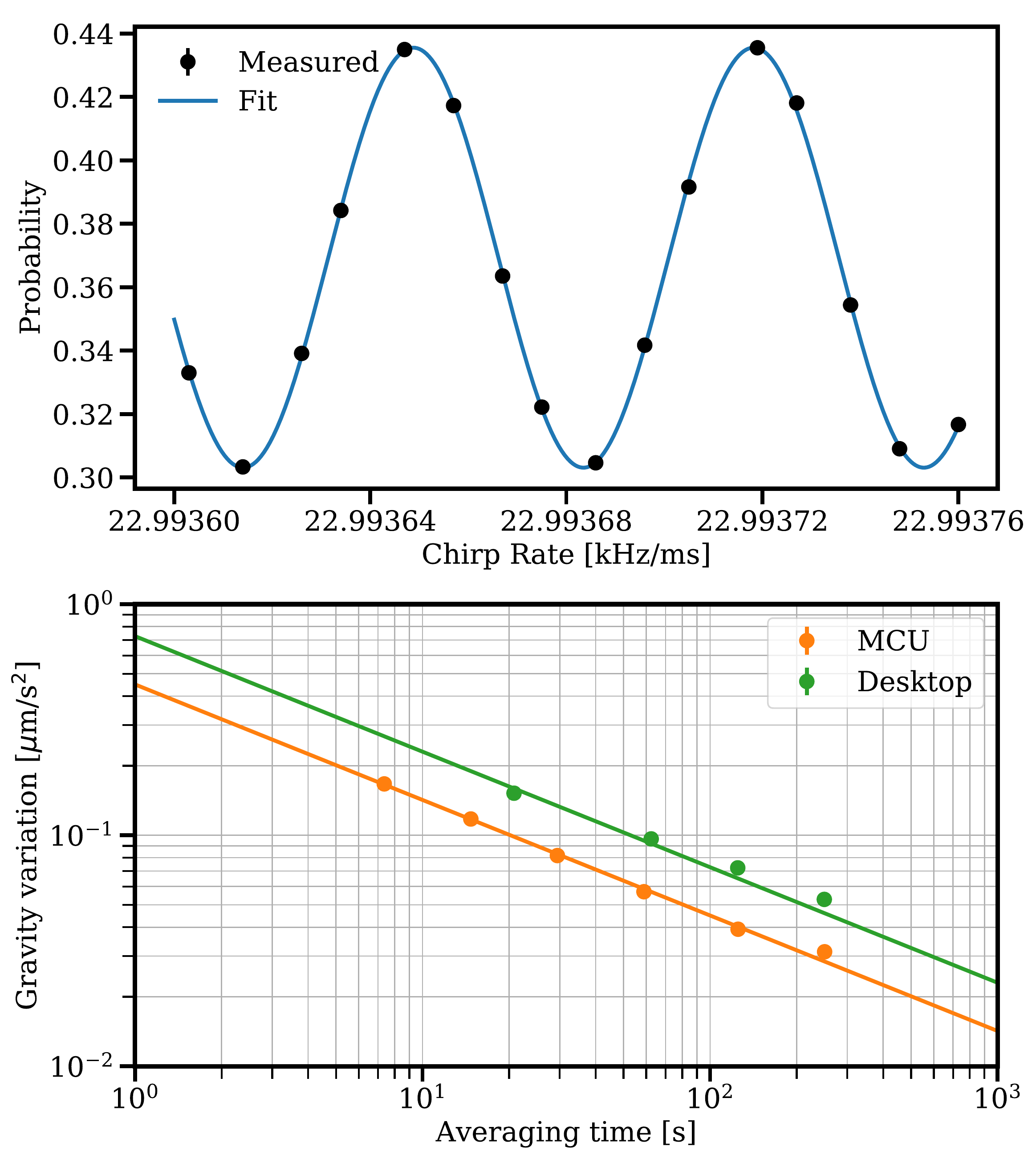}
    \caption{Top: averaged data and fit for 6529 fringes; the error bars are smaller than the point size. Bottom: Comparison of Allan deviations between the MCU and the desktop-based system due to cycle time improvements.}
    \label{fig:results}
\end{figure}

The embedded control system successfully operated all aspects of the atom interferometer, measured gravity, and improved the data rate. A timing sequence of select signals for measuring gravity is shown in figure \ref{fig:timing}. To measure gravity, this sequence is repeated while varying the frequency sweep rate in the Doppler-sensitive pulses. Results for measuring gravity are shown in figure \ref{fig:results}. The fluorescence detection measures the ratio of atom in one of the states relative to the total number of atoms. The population of atoms in each state is computed on the MCU. The value of gravity is extracted from the total phase of the interferometer\cite{wu2017multiaxis}. Sensitivity to gravity variations is determined by taking successive measurements then computing an Allan deviation. An improved data rate is reflected in the improved sensitivity for an equivalent measurement time.

The data used for the Allan deviation are 16-point fringes, as seen aggregated in figure \ref{fig:results}. The MCU is faster due to onboard integration and eliminated dead time. A single fringe point requires only 460 ms, more than a factor of 2 improvement compared to 1.3 s for the desktop-based system. As shown, the sensitivity for a single measurement using the MCU improved by more than a factor $\sqrt{2}$, at 450 nm/s$^2$/$\sqrt{\text{Hz}}$, compared to the previous desktop computer-based system, at 727 nm/s$^2$/$\sqrt{\text{Hz}}$.

\section{Conclusion}
This embedded control system successfully operated the mobile atomic sensor and eliminated unnecessary dead time resulting more sensitive measurements. This MCU-based system delivers comparable performance to existing systems, can be operated in a power-constrained environment, and provides the flexibility of real-time data processing and decision making. This simple architecture can be extended to any other atomic sensor and has proven to be a viable alternative to controllers based on desktop computers and FPGAs. This system paves the way for portable, efficient, and flexible control of atomic sensors as they proliferate in field applications.


\begin{acknowledgments}
We thank Eric Copenhaver, Matt Jaffe, and Timothy Nguyen for their support and productive discussions. Funding Sources: Bakar Fellows Program; NASA Planetary Instrument Definition and Development Program through a Contract with Jet Propulsion Laboratory (JPL) award number 1465360.
\end{acknowledgments}

\bibliography{references.bib}

\end{document}